# Size Distribution of Superparamagnetic Particles Determined by Magnetic Sedimentation

J.-F. Berret@, O. Sandre and A. Mauger

*Matière et Systèmes Complexes, UMR 7057 CNRS Université Denis Diderot Paris-VII, 140 rue de Lourmel, F-75015 Paris France - Laboratoire Liquides Ioniques et Interfaces Chargées, UMR 7612 CNRS Université Pierre et Marie Curie Paris-VI, 4 place Jussieu, F-75252 Paris Cedex 05 France - Département Mathématique Informatique Physique Planète et Univers, CNRS, 140 rue de Lourmel, F-75015 Paris France*

**Abstract :** We report on the use of magnetic sedimentation as a means to determine the size distribution of dispersed magnetic particles. The particles investigated here are *i)* single anionic and cationic nanoparticles of diameter D ~ 7 nm and *ii)* nanoparticle clusters resulting from electrostatic complexation with polyelectrolytes and polyelectrolyte-neutral copolymers. A theoretical expression of the sedimentation concentration profiles at the steady state is proposed and it is found to describe accurately the experimental data. When compared to dynamic light scattering, vibrating sample magnetometry and cryogenic transmission electron microscopy, magnetic sedimentation exhibits a unique property : it provides the core size and core size distribution of nanoparticle aggregates.

## I – Introduction

Magnetic nanoparticles are currently used in a wide variety of material science and biomedical applications[1,2]. Important technological advances have been achieved in the purification of biomolecules and in cell separation techniques. Surface-modified nanoparticles have been also developed for Magnetic Resonance Imaging (MRI) and drug delivery. In this context, the functionalisation of nanoparticles has attracted in recent years much attention because it allows the formation of nanostructures of well-defined physical properties.

Among the different inorganic molecules and macromolecules used to modify the surface of magnetic nanoparticles, polymers have been one of the most investigated[3-16]. Polymers are of interest because they can form a diffuse and neutral shell around the particles and therefore increase the colloidal stability of the hybrid system. Dextran is a polysaccharide which is already used commercially as coating agent for magnetic nanoparticles. As a result, dextran-coated nanoparticles were studied extensively in the last years[2-4,7]. The interactions between the glucose units of dextran and the surface charges of the particles are weak and more likely of hydrogen bond type. Recently, novel polymer structures such as polyelectrolytes[6,17], dendrimers[8] and block copolymers[9-12,16] have emerged as alternative and appealing coating systems. In polymer-nanoparticle hybrids, the question of the microstructure of the mixed colloids remains crucial since it determines the properties that are relevant for applications. In such cases, it is important to know the state of aggregation of nanoparticles, as well as the proportions and sizes of the organic *versus* inorganic species. In the context of hierarchical microstructures, dynamic light scattering which provides the value of an equivalent hydrodynamic diameter is notably insufficient.

Magnetic sedimentation and magnetophoresis have been used for separation purposes, generally in combination with magnetically loaded colloids or cells in the micrometer range[18-25]. In the present paper, we show that magnetic sedimentation can be employed successfully as a means to derive the size distribution of magnetic particles in the nanometer range. For uncoated nanoparticles, the magnetic sedimentation results compare well with those obtained by vibrating sample magnetometry and cryogenic transmission microscopy (Cryo-TEM). For polymer-coated magnetic clusters, magnetic sedimentation allow to estimate the average aggregation number.

## II - Theory

### *Monodisperse Magnetic Particles*

The force exerted on a magnetic nanoparticle dispersed in a solvent and submitted to a magnetic field gradient reads[26] :

$$\mathbf{F_{Mag}} = \mu_0 V (\mathbf{M} \cdot \nabla) \mathbf{H} \qquad (1)$$

where V is the volume of the nanoparticle, M its magnetization in a given field H and $\mu_0$ the permeability in vacuum. In one-dimension, *i.e.* for a field gradient along the z-direction, and for superparamagnetic particles with volumetric magnetization $m_S$, the magnetic force expresses as :

$$F^z_{Mag} = \mu_0 \frac{\pi D^3}{6} m_S L(\zeta(H)) \frac{dH(z)}{dz}. \qquad (2)$$

In Eq. 2, D denotes the diameter of the particle, $\zeta(H) = \mu_0 \pi D^3 m_S H / 6 k_B T$ and $L(\zeta) = \coth \zeta - 1/\zeta$ is the Langevin function. In the following, we study the effect of a constant magnetic field gradient dH(z)/dz on a ferrofluid solution of height h and of number density of particle $n_0$. $n_0$ is defined as the number of particles per unit volume, expressed in $cm^{-3}$. Before the sedimentation takes place, the number density of particles is uniform in the region of space occupied by the dispersion, 0 < z < h. In a magnetic field gradient, the number density evolves with time and altitude. We define by n(t,z) the number density of particles at the altitude z at a time t > 0 of the sedimentation process. The partial differential equation for n(t,z) is given by the Fokker-Planck equation[27,28]:

$$\frac{\partial n(t,z)}{\partial t} = D_0 \frac{\partial^2 n(t,z)}{\partial z^2} - \mu F_{Mag}^z \frac{\partial n(t,z)}{\partial z} \quad (3)$$

where μ the mobility given by the Stokes formula, μ = 1/(3πηD) and η is the viscosity of the solvent. Here, we also assume that the particles are at all time in thermal equilibrium with respect to their velocity distribution, implying that the diffusion coefficient $D_0$ obeys the Nernst-Einstein equation $D_0 = \mu k_B T$. For a constant magnetic field gradient, the steady state solution $n_S(z)$ of the partial differential equation reads :

$$n_S(z) = h \frac{n_0}{\lambda(1 - \exp[-h/\lambda])} \exp[-z/\lambda] \quad (4)$$

Eq. 4 introduces the length scale λ characteristic of the magnetic sedimentation process. λ is defined as the ratio between the thermal energy and the magnetic force exerted on a single particle.

$$\lambda = \frac{6k_B T}{\mu_0 \pi D^3 m_S L(\zeta(H)) \left(\frac{dH}{dz}\right)} \quad (5)$$

It is interesting to note in Eq. 3 that the stationary state solution $n_S(z)$ does not depend on the hydrodynamic diameter (or on the mobility) of the particles. This result implies that the number density profile shown by Eq. 4 is only a function of the total magnetic load of the particles, and not of the polymer coating that is attached to it. For the dilute dispersions investigated in this work, the nanoparticle volume fraction $\phi = \pi n_0 D^3 / 6$ is low and as a result, the magnetization of the fluid remains much smaller than the applied excitation field H. The magnetic induction within the sample then reduces to B ≈ $\mu_0$H. Under these conditions, Eq. 5 rewrites :

$$\lambda = \frac{6k_B T}{\pi D^3 m_S L(\zeta(B)) \left(\frac{dB}{dz}\right)} \quad (6)$$

where $\zeta(B) = \pi D^3 m_S B / 6k_B T$. With monodisperse (D = 10 nm) superparamagnetic ($m_S$ = 2.6×10$^5$ A m$^{-1}$) particles submitted to a magnetic field of 0.2 T and to a magnetic field gradient of 20 T m$^{-1}$, one gets λ = 1.8 mm.

*Polydisperse Magnetic Particles*
In order to describe magnetic colloidal dispersions subjected to sedimentation, the polydispersity of the particles has to be taken into account. We assume for the particle diameters a probability distribution function of log-normal type[29,30] :

$$p(D, \tilde{D}, s) = \frac{1}{\sqrt{2\pi}\beta(s)D} \exp\left(-\frac{\ln^2(D/\tilde{D})}{2\beta(s)^2}\right) \quad (7)$$

where $\tilde{D}$ is defined as the median diameter and β(s) is related to the polydispersity index s by the relationship $\beta(s) = \sqrt{\ln(1+s^2)}$. The polydispersity index is defined as the ratio between the standard deviation and the average diameter[30]. For dilute solutions, *i.e.* for weight concentrations below 1 wt. %, the magnetic and colloidal interactions between particles can be neglected[31]. In such a case, the stationary sedimentation profiles obtained for different sizes of particles are additive and express as :

$$n_S(z, \tilde{D}, s) = n_0 h \int_0^\infty \frac{\exp[-z/\lambda]}{\lambda(1 - \exp[-h/\lambda])} p(D, \tilde{D}, s) dD \quad (8)$$

Within these approximations, the number density $n_S(z, \tilde{D}, s)$ becomes also a function of $\tilde{D}$ and s, as specified in Eq. 8. In order to allow comparison with experiments, the number density in Eq. 8 has to be rewritten in terms of the weight concentration of magnetic particles $c_S(z, \tilde{D}, s)$ :

$$\frac{c_S(z, \tilde{D}, s)}{c_0} = \frac{h}{\overline{D^3}} \int_0^\infty \frac{D^3 \exp(-z/\lambda)}{\lambda(1 - \exp(-h/\lambda))} p(D, \tilde{D}, s) dD \quad (9)$$

Here, $\overline{D^3}$ is the third moment of the particle size distribution. Fig. 1 displays the concentration profiles $c_S(z, \tilde{D}, s)/c_0$ predicted for 10 nm nanoparticles with different polydispersity s = 0.1 – 0.4. The parameters used for the simulations are specified in the captions. In this semilogarithmic representation, the straight line found for s ≤ 0.1 with a slope 1/λ accounts for the exponential decrease predicted by Eq. 4. Moreover, due to the strong dependence of λ with the particle diameter (λ ~ D$^{-3}$), the concentration profile of the magnetic solution is very sensitive to the polydispersity. In the following, we use this property to determine experimentally the size distribution of magnetic dispersions.

## III - Experimental

Superparamagnetic nanoparticles of maghemite (γ-Fe$_2$O$_3$) were synthesized by alkaline co-precipitation of iron II and iron III salts. Three batches, noted S1, S2 and S3 in the following have been studied with respect to complexation with polymers. The batches S1 and S2 were sorted in sizes by successive phase separations, according to protocols described in Ref. 32. For S1 and S3, the particles were coated with citrate ligands and thus negatively charged at neutral pH, whereas S2 was prepared in acidic conditions (pH 1.9) with nitrate counterions adsorbed on their surfaces.

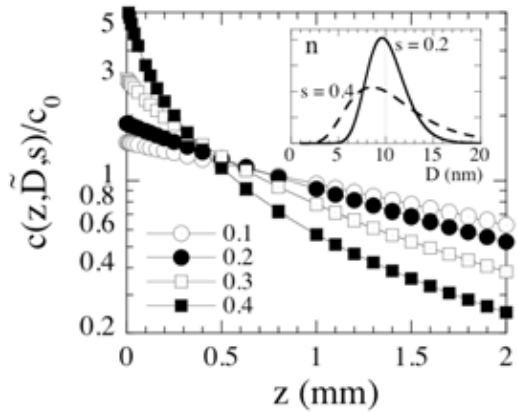

**Figure 1** : Concentration profiles $c_S(z,\tilde{D},s)/c_0$ calculated from Eq. 9 for a 10 nm magnetic nanoparticle dispersion of different polydispersity $s = 0.1 – 0.4$. Experimental parameters are $h = 2$ mm, $m_S = 3\times10^5$ A m$^{-1}$ and $dB/dz = 10$ T m$^{-1}$. The coefficient $L(\zeta(B))$ in Eq. 7 is set to unity. The concentration profiles obtained by sedimentation are found to be strongly dependent on the polydispersity of the particles.

At the concentrations at which the synthesis were made, c ~ 5 wt. %, these magnetic dispersions are thermodynamically stable over a period of several years. The magnetic nanosols were characterized by different techniques, which included electron microdiffraction, dynamic light scattering, cryo-transmission electron microscopy, vibrating sample magnetometry and magnetic sedimentation. The set-ups and protocols for light scattering, cryo-TEM and magnetometry have been described in Refs. 16, 33 and 34 and we refer to this work for further details. Electron microdiffraction spectra of the $\gamma$-Fe$_2$O$_3$ spinel structure realized on batch S1 are provided in the supporting information section. Also displayed are the list and assignment of the Bragg reflections relative to this structure[35].

For the present study, electrostatic complexation was performed with three polymer architectures. The first one is a poly(acrylic acid) with molecular weight 2000 g mol$^{-1}$, hereafter noted PAA$_{2K}$. It was purchased from Fluka (ref. 81130) and used as is. The second polymer is a poly(acrylic acid)-*graft*-poly(ethylene oxide) comb polymer, or PAA$_{4K}$-*g*-PEO[36]. This polymer was kindly provided to us by Rhodia. Along the PAA backbone of molecular weight 4000 g mol$^{-1}$, PEO segments (molecular weight 1000 g mol$^{-1}$) are grafted randomly, yielding a total average molecular weight for the comb of 29000 g mol$^{-1}$ with a polydispersity of 1.9. Titration experiments performed on solutions containing these weak polyacids have confirmed the values of the molecular weights for the charged segments. The third polymer investigated is a cationic-neutral diblock copolymer, referred to as poly(trimethylammonium ethylacrylate)-*b*-poly(acrylamide) and abbreviated as PTEA-*b*-PAM in the text below[37]. The polyelectrolyte block (PTEA) is a strong polyelectrolyte and as such its monomers are fully ionized. The monomers are positively charged at neutral pH. In the present study, we utilized two molecular weights for the cationic block, 5000 g mol$^{-1}$ and 11000 g mol$^{-1}$. For both the poly(acrylamide) block was 30000 g mol$^{-1}$. The polydispersity index of the two diblocks was estimated by size exclusion chromatography at 1.6. The pH of the polymer and nanoparticle suspensions was adjusted with reagent-grade nitric acid (HNO$_3$) and with sodium hydroxide (NaOH).

Magnetic sedimentation experiments were carried out at $\gamma$-Fe$_2$O$_3$ concentrations c = 0.1 – 0.5 wt. %. , *i.e.* for volume fraction $\phi$ = 0.02 – 0.1 %. At such low $\phi$, we insure that the colloidal and magnetic interactions are weak[31] and that the particles can be described as non interacting Brownian particles, as in Eq. 3. The solutions were placed in 1 or 4 mm thick Hellma cells above a permanent ferrite or rare-earth magnet. The magnetic field above the magnet was measured using a GM05 Gaussmeter from Hirst Magnetic Instruments as a function of the distance z, and thereafter the gradient was computed.

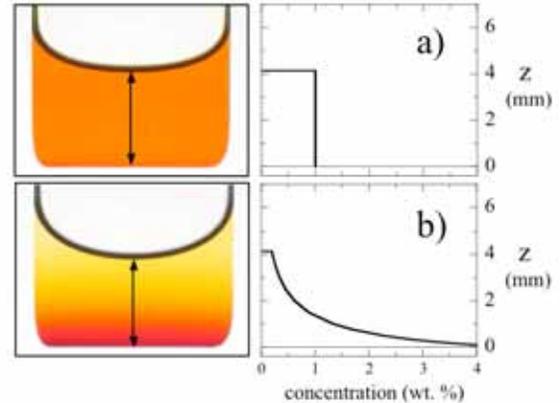

**Figure 2** : Photographs and concentration profiles of a magnetic dispersion at the initial (a) and final (b) stages of the sedimentation process. The solution shown here is from batch S2 at a concentration c = 1 wt. %. The magnet is located at the bottom of the cell and the field gradient (40 ± 4 T m$^{-1}$) is constant over the height of the sample.

Depending on the configuration, gradients between 1 and 40 T m$^{-1}$ were obtained. In the experiments, it was verified that the gradients $(dB/dz)$ were constant and homogeneous over the entire volume of the solution. The variations of the Langevin function $L(\zeta(B))$ in Eq. 6 were also found to be weak with respect to the altitude z. The iron oxide concentrations $c_S(z)$ were determined from images of the solutions taken at different times of the settling process with a G5 Canon camera and subsequently by analysis based on colorimetry. Images and concentration profiles of a S2-dispersion at c = 1 wt. % are shown in Fig. 2 at the initial and final stage of sedimentation. The kinetics of sedimentation will be shown in a forthcoming paper. The spatial resolution in altitude z is of the order of 20 µm.

## IV – Results and Discussion

### IV. 1 – Single nanoparticles

Vibrating sample magnetometry has become an increasingly important tool for the determination of the particle size distribution of magnetic dispersions[29,32,38]. The experiment consisted in measuring the magnetization *versus* excitation curve M(H) for a solution at concentration $c_0$. The bottom curve in Fig. 3a shows the evolution of the macroscopic magnetization M(H) normalized by its saturation value $M_{Sat}$ for the S1-batch.

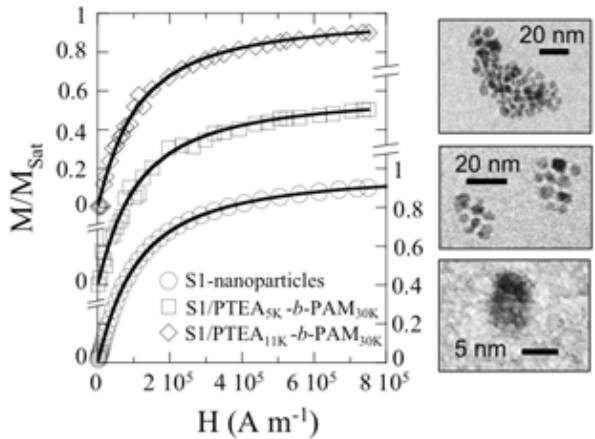

*Figure 3a* : Magnetic field dependences of the macroscopic magnetization M(H) normalized by its saturation value $M_{Sat}$ for S1 (circles), S1/PTEA$_{5K}$-b-PAM$_{30K}$ (squares) and S1/PTEA$_{11K}$-b-PAM$_{30K}$ (lozenges) dilute solutions. The solid curves were obtained using the Langevin function for paramagnetism convoluted with a log-normal distribution function for the particle sizes. The results of best fit calculations are the median diameter $\tilde{D}_{Mag}$ and the polydispersity of $s_{Mag}$ (Table I). On the right hand side in Fig. 3a, cryo-TEM images of single nanoparticles (lower inset, S1-batch) and of polymer-nanoparticle aggregates illustrate the microstructure of the colloids investigated in this work. The middle and top insets are from S1/PTEA$_{5K}$-b-PAM$_{30K}$ and S1/PTEA$_{11K}$-b-PAM$_{30K}$ complexes, respectively[34].

With the notations of Eq. 2, $M_{Sat} = \phi m_S$, where $\phi = \pi n_0 \overline{D^3}/6$ denotes the volume fraction of particles in the solution and $m_S$ the volumetric magnetization of maghemite ($m_S$ = 2.6×10$^5$ A m$^{-1}$). The solid curve through the data points was obtained using the Langevin function for paramagnetism convoluted with a log-normal distribution function of the particle size[26,29].

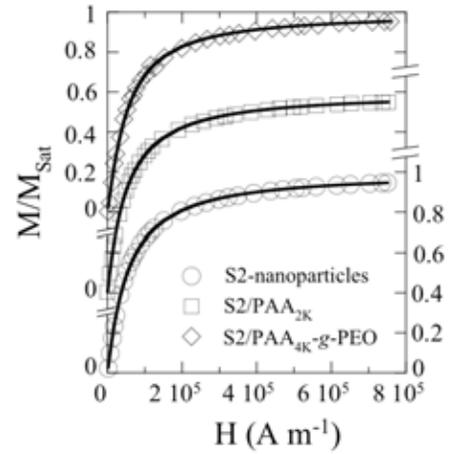

**Figure 3b** : Same as in Fig. 3a for S2, S2/PAA$_{2K}$ and S2/PAA$_{4K}$-g-PEO dilute solutions respectively.

The parameters of the distribution (Eq. 6) are the median diameter $\tilde{D}_{Mag}$ = 6.3 ± 0.3 nm and the polydispersity of $s_{Mag}$ = 0.23 ± 0.03, where the index "Mag" refers to magnetometry. On the right hand side of Fig. 3, a cryo-TEM image of a single nanoparticle is shown as an illustration. It confirms that the particles are in the nanometer range and of spherical symmetry. The image analysis of 470 particles (S1-batch) captured by cryo-TEM provided an additional determination of the particle size distribution[16]. The distribution was found to be well accounted for by a log-normal function with a median diameter $\tilde{D}_{TEM}$ = 6.3 ± 0.2 nm and a polydispersity $s_{TEM}$ = 0.27 ± 0.04.

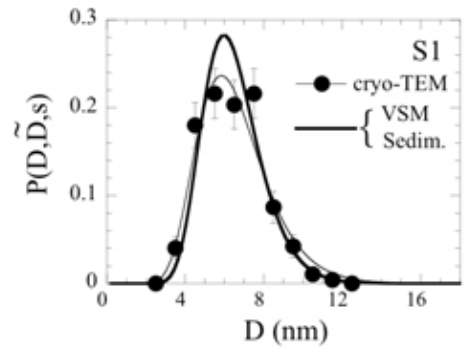

**Figure 4** : Size distribution $p(D, \tilde{D}, s)$ resulting from an image analysis of 470 citrate-coated particles (cryo-TEM, batch S1). The data points were fitted using a log-normal function (continuous thin line, Eq. 7) with $\tilde{D}_{TEM}$ = 6.3 ± 0.2 nm and $s_{TEM}$ = 0.27 ± 0.04. The cryo-TEM data are compared with the distributions received from vibrating sample magnetometry (VSM) and magnetic sedimentation (continuous thick line).

The size distributions obtained by vibrating sample magnetometry and cryo-TEM are compared in Fig. 4, and within the experimental accuracies they are in excellent agreement with each other. Vibrating sample magnetometry carried out on S2 (lower curve in Fig.

3b) and S3 solutions (data not shown) yields $\tilde{D}_{Mag}$ = 7.1 nm and $s_{Mag}$ = 0.26, and $\tilde{D}_{Mag}$ = 7.0 nm and $s_{Mag}$ = 0.36, respectively. As for S1, the relative uncertainties are 5 % on the diameter and 10 % on the polydispersity. The results obtained by magnetometry for the uncoated nanoparticles are summarized in Table I. These values are also compared with the hydrodynamic diameters $D_H$ received from light scattering experiments. For samples S1 to S3, the polydispersity s varies from 0.23 to 0.36 and $D_H$ is found to increase from 11 nm to 27 nm. These findings confirm the strong variation of the hydrodynamic properties of colloids with respect to the size polydispersity[30].

Figs. 5a, 5b and 5c displays the concentration profiles $c_S(z, \tilde{D}, s)/c_0$ for the three nanoparticles batches S1 - S3, using magnetic field gradient of 28, 18 and 18 T m$^{-1}$ respectively. The data are shown with the same scales to emphasize the effect of the polydispersity on the concentration. As suggested by Fig. 1, with increasing s the local concentration is strongly increased in the vicinity of the magnet (corresponding to z = 0). The solid curves through the data are best fit calculations using Eq. 9, with adjustable parameters $\tilde{D}_{Sed}$ (median diameter) and $s_{Sed}$ (polydispersity). Here and below, the index "Sed" refers to the sedimentation experiments. In the three solutions, the agreement between the data and the calculations is excellent. In addition, the $\tilde{D}_{Sed}$ and $s_{Sed}$ values listed in Table I confirmed those found by the vibrating sample magnetometry. It should be mentioned that for the calculated profiles, the parameter $L(\zeta(B))$ in Eq. 7 has been set to unity, instead of its actual value, 0.75. This approximation is discussed in the conclusion section. The good agreement found between the two techniques demonstrates the reliability of magnetic sedimentation to derive the size distribution of magnetic nanoparticles.

### IV. 2 – Electrostatic complexation between polymers and nanoparticles

In order to adsorb ion-containing polymers to the surface of the nanoparticles, we have followed protocols that were described recently in the literature[16,39,40]. For the PAA-based polymers (*i.e.* for PAA$_{2K}$ and PAA$_{4K}$-*g*-PEO), we exploit the precipitation-redispersion mechanism which was first evidenced on 7 nm cerium oxide nanoparticles[17]. The precipitation of the cationic $\gamma$-Fe$_2$O$_3$ dispersion (batch S2) by oppositely charged polyelectrolytes is performed in acidic conditions. As the pH of the solution is increased by addition of sodium hydroxide, the precipitate redisperses spontaneously, yielding a clear solution that now contains polymer-nanoparticle hybrids. As shown in Table I, the hydrodynamic sizes of the S2/PAA$_{2K}$ and S2/PAA$_{4K}$-*g*-PEO systems are $D_H$ = 22 and 30 nm, respectively. These values are 7 and 15 nm larger than the hydrodynamic diameter of the uncoated particles. This increase could be due either to the presence of a polymer brush surrounding the particles, to an increase of the polydispersity, or to the formation of doublets, triplets etc… of nanoparticles. As shown below, magnetic sedimentation allows us to discriminate between these different assumptions.

|  | specimens | light scattering $D_H$ (nm) | vibrating sample magnetometry $\tilde{D}_{Mag}$ (nm) | $s_{Mag}$ | magnetic sedimentation $\tilde{D}_{Sed}$ (nm) | $s_{Sed}$ |
|---|---|---|---|---|---|---|
| nanoparticle | S1 (anionic) | 11 | 6.3 | 0.23 | 6.3 | 0.23 |
| nanoparticle | S2 (cationic) | 15 | 7.1 | 0.26 | 7.3 | 0.30 |
| nanoparticle | S3 (anionic) | 27 | 7.0 | 0.36 | 7.0 | 0.40 |
| Hybrid colloids | S2/PAA$_{2K}$ | 22 | 7.3 | 0.26 | 8.0 | 0.24 |
| Hybrid colloids | S2/ PAA$_{4K}$-*g*-PEO | 30 | 7.7 | 0.21 | 9.0 | 0.29 |
| Hybrid colloids | S1/PTEA$_{5K}$-*b*-PAM$_{30K}$ | 92 | 6.0 | 0.23 | 23 | 0.18 |
| Hybrid colloids | S1/PTEA$_{11K}$-*b*-PAM$_{30K}$ | 172 | 6.0 | 0.22 | n.d. | n.d. |

*Table I :* Particle diameters determined by dynamic light scattering ($D_H$), vibrating sample magnetometry ($\tilde{D}_{Mag}$) and magnetic sedimentation ($\tilde{D}_{Sed}$) for magnetic nanoparticle and polymer-nanoparticles hybrids. For magnetometry and sedimentation, the polydispersity index s is also obtained. Typical uncertainties for these two techniques are 5 % on the diameter and 10 % on the polydispersity. Cryo-TEM experiments performed of S1-particles resulted in a median diameter $\tilde{D}_{TEM}$ = 6.3 ± 0.2 nm and a polydispersity s = 0.27 ± 0.04 (Fig. 4). For the bare of citrate-coated nanoparticles, the agreement between magnetometry, cryo-TEM and sedimentation is excellent. For particles clusters, the magnetic sedimentation results in an equivalent magnetic diameter. Cryo-TEM data on S1/PTEA$_{5K}$-*b*-PAM$_{30K}$ complexes yielded $\tilde{D}_{TEM}$ = 20.5 ± 1 nm and s = 0.18 ± 0.03 (Fig. 6).

The second protocol deals with a collective clustering of the anionic citrate-coated nanoparticles driven by the addition of oppositely charged copolymers[16,39,40]. Polymer-nanoparticle complexes were obtained by mixing stock solutions prepared at the same weight concentration and pH (pH 8). The solutions investigated here were prepared at the preferred mixing ratio *i.e.* at the ratio where all the components (polymers and nanoparticles) associate to form colloidal complexes[41]. With citrate coated nanoparticles (S1), the preferred mixing ratio is 1 for PTEA$_{5K}$-*b*-PAM$_{30K}$ and 2 for PTEA$_{11K}$-*b*-PAM$_{30K}$. The structure of the colloidal complexes obtained by collective clustering process has been disclosed recently[16]. It was demonstrated by a combination of light scattering and cryo-TEM that the mixed aggregates have a core-shell microstructure. In the core, the polyelectrolyte blocks and the oppositely charged nanoparticles are tightly bound together, forming a dense magnetic cluster in the range 20 - 100 nm.

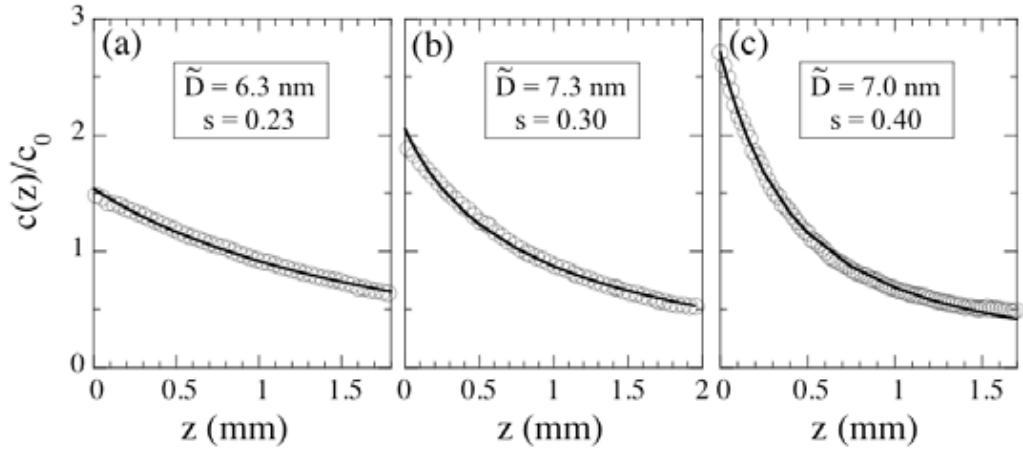

**Figure 5**: Concentration profiles $c_S(z,\tilde{D},s)/c_0$ for S1 (a), S2 (b) and S3 (c) dilute solutions. The magnetic field gradients used in these experiments are of 28, 18 and 18 T m$^{-1}$ respectively. The solid curves are best fit calculations using Eq. 9 and adjustable parameters $\tilde{D}_{Sed}$ and $s_{Sed}$. The values for $\tilde{D}_{Sed}$ and $s_{Sed}$ are given in Table 1.

Cryo-TEM images of cores are shown in Fig. 3a. The middle and top insets on the right hand side of Fig. 3a are from S1/PTEA$_{5K}$-b-PAM$_{30K}$ and S1/PTEA$_{11K}$-b-PAM$_{30K}$ complexes, respectively[34]. As notified recently[16], the average distance between particle comprised in a cluster has been estimated at 8.1 ± 0.1 nm, i.e. slightly larger than the diameter of a particle ($\tilde{D}$ = 6.3 nm). The value of 8.1 nm corresponds to a magnetic volume fraction of 0.32 ± 0.03 in the clusters[42]. For the analysis of the cluster morphology (sample S1/PTEA$_{5K}$-b-PAM$_{30K}$), it was assumed that the 2D-projections of the clusters could be represented by ellipses with major and minor axis noted *a* and *b* respectively. Based on the image analysis of 200 aggregates, the probability distribution functions for the minor and major axis, as well as for the equivalent diameter D$_{TEM}$ = $(ab)^{1/2}$ were obtained. For S1/PTEA$_{5K}$-b-PAM$_{30K}$ mixed solutions, the clusters were found to be slightly anisotropic[16,34]. Fig. 6 displays the size distribution of the S1/PTEA$_{5K}$-b-PAM$_{30K}$ clusters as received from cryo-TEM. The distribution was found to obey a log-normal function with an median diameter $\tilde{D}_{TEM}$ = 20.5 ± 1 nm and a polydispersity $s_{TEM}$ = 0.18 ± 0.03 (thick continuous line in Fig. 6). For these samples, dynamic light scattering have also revealed the presence of a neutral shell around the cores. The hydrodynamic diameters of the hybrid colloids were found at D$_H$ = 92 nm and D$_H$ = 172 nm for mixed colloids prepared respectively with PTEA$_{5K}$-*b*-PAM$_{30K}$ and PTEA$_{11K}$-*b*-PAM$_{30K}$ (Table I). Note that due to different mixing conditions, D$_H$ for S1/PTEA$_{5K}$-*b*-PAM$_{30K}$ hybrids is slightly larger than that reported recently[16].

Vibrating sample magnetometry have been performed on dilute solutions containing the four types of mixed systems mentioned previously, namely S2/PAA$_{2K}$, S2/PAA$_{4K}$-*g*-PEO, S1/PTEA$_{5K}$-*b*-PAM$_{30K}$ and S1/PTEA$_{11K}$-*b*-PAM$_{30K}$. The magnetometry results are shown by the two top curves of Figs. 3a and 3b, together with best fits calculations using the superparamagnetic Langevin function convoluted with a log-normal size distribution[29]. Although the magnetic nanoparticles are now associated with polymers, or in some cases arranged into large and tight clusters, the reduced magnetization M/M$_{Sat}$ is similar to that of the bare particles.

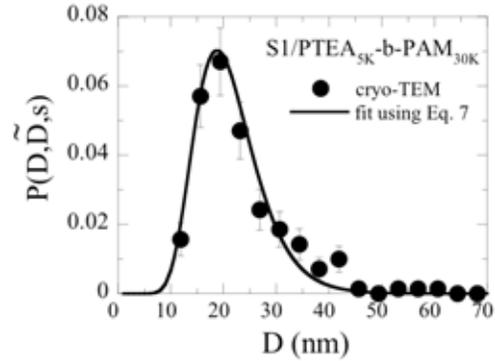

**Figure 6**: Size distribution $p(D,\tilde{D},s)$ obtained by cryo-TEM on 200 S1/PTEA$_{5K}$-b-PAM$_{30K}$ clusters. The data points were fitted using a log-normal function (continuous thick line, Eq. 7) with a median diameter $\tilde{D}_{TEM}$ = 20.5 ± 1 nm and a polydispersity $s_{TEM}$ = 0.18 ± 0.03. These data are in good agreement with those of magnetic sedimentation performed on the same system.

In Fig. 3a, the continuous lines were obtained using $\tilde{D}_{Mag}$ = 6.0 nm, $s_{Mag}$ = 0.23 (for S1/PTEA$_{5K}$-*b*-PAM$_{30K}$) and $\tilde{D}_{Mag}$ = 6.0 nm, $s_{Mag}$ = 0.22 (for S1/PTEA$_{11K}$-*b*-PAM$_{30K}$), whereas in Fig. 3b the values are $\tilde{D}_{Mag}$ = 7.3, $s_{Mag}$ = 0.26 (for S2/PAA$_{2K}$), and $\tilde{D}_{Mag}$ = 7.7 nm, $s_{Mag}$ = 0.21 (for S2/PAA$_{4K}$-*g*-PEO). These later findings suggest that for magnetic hybrid colloids submitted to a homogeneous and constant magnetic field (as in magnetometry), the moments associated to the nanoparticles are not adding their contributions. An estimate of the magnetic dipolar interaction between 6.3 nm particles separated by a particle diameter provides an energy $\pi\mu_0 m_S^2 \overline{D^3}/144 = 0.11 k_B T^{26}$. As a result, the moments appear uncorrelated in these experiments and vibrating sample magnetometry does not allow the determination of the cluster sizes.

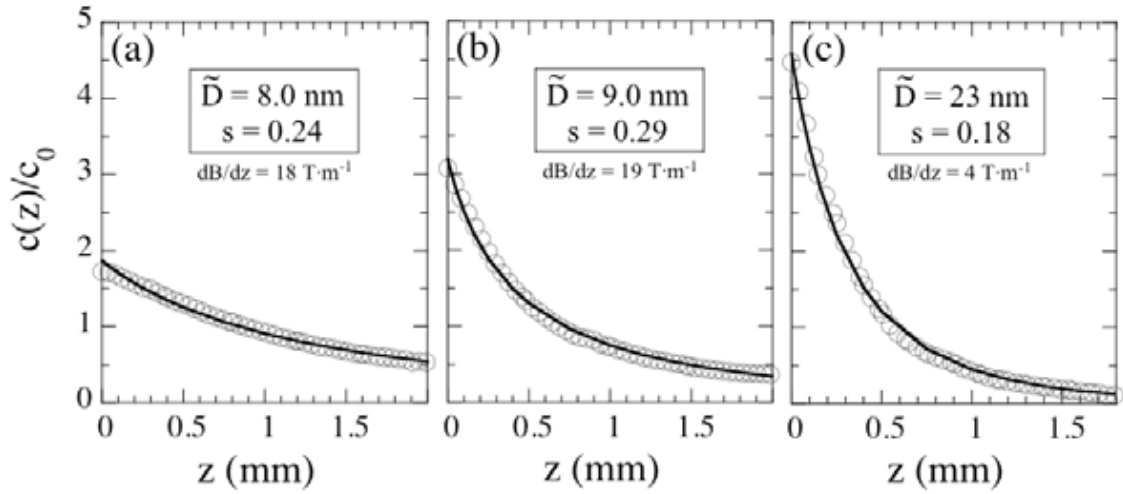

**Figure 7 :** Concentration profiles $c_S(z,\tilde{D},s)/c_0$ for S2/PAA$_{2K}$ (a), S2/PAA$_{4K}$-g-PEO (b) and S1/PTEA$_{5K}$-b-PAM$_{30K}$ (c) dilute solutions. The magnetic field gradients used in these experiments are 18, 19 and 4 T m$^{-1}$, respectively. The solid curves are from Eq. 9, using for adjustable parameters $\tilde{D}_{Sed}$ and $s_{Sed}$ (see Table 1).

Magnetic sedimentation was performed on S2/PAA$_{2K}$, S2/PAA$_{4K}$-g-PEO and S1/PTEA$_{5K}$-b-PAM$_{30K}$ complexes, again in dilute solutions. Figs. 7a, 7b and 7c shows the concentration profiles $c_S(z,\tilde{D},s)/c_0$ for these systems, using magnetic field gradients of 18, 19 and 4 T m$^{-1}$, respectively. The data are shown using the same ordinate scale in order to emphasize the effect of the clustering on the sedimentation profiles. Note that in these experiments S1/PTEA$_{5K}$-b-PAM$_{30K}$ has been subjected to a magnetic field gradient that is 5 times lower than the two other samples. The solid curves through the data are best fit calculations using Eq. 9, with adjustable parameters $\tilde{D}_{Sed}$ and $s_{Sed}$. We found $\tilde{D}_{Sed}$ = 8.0 nm, $s_{Sed}$ = 0.24 (S2/PAA$_{2K}$), $\tilde{D}_{Sed}$ = 9.0 nm, $s_{Sed}$ = 0.29 (S2/PAA$_{4K}$-g-PEO), and $\tilde{D}_{Sed}$ = 23 nm, $s_{Sed}$ = 0.18 (S1/PTEA$_{5K}$-b-PAM$_{30K}$). The values for the later system are in qualitative agreement with those received from the cryo-TEM experiments (Fig. 6). These findings suggest that for magnetic hybrid colloids submitted to a magnetic field gradient, the moments borne by the nanoparticles are additive and as such the clusters are associated with large equivalent magnetic moment. In this case, magnetic sedimentation remains sensitive to both size and size distribution of the clusters.

From the above distributions, the average aggregation numbers of the hybrid colloids were estimated. For a cluster made from N particles, the diameter $D_{Sed}(N)$ observed by magnetic sedimentation expresses as a function of the single particle diameter $D_{Sed,N=1}$ as :

$$D_{Sed}^3(N) = N D_{Sed,N=1}^3. \quad (10)$$

Combining the data in Table I with Eq. 10, the average aggregation numbers, noted $N_{Ave}$ were calculated. We found $N_{Ave}$ = 1.14, 1.82 and 45 for S2/PAA$_{2K}$, S2/PAA$_{4K}$-g-PEO and S1/PTEA$_{5K}$-b-PAM$_{30K}$, respectively. From these estimates, we conclude that for S2/PAA$_{2K}$, the adsorption of the PAA$_{2K}$-coating has not modified the dispersion state of the particles. A similar result was found for the nanoceria[17]. For PAA$_{4K}$-g-PEO, there is probably a slight aggregation of the nanoparticles during the precipitation-redispersion process, which could take the form of doublets, or triplets of particles. For S1/PTEA$_{5K}$-b-PAM$_{30K}$ clusters, we confirm the self-assembly into large clusters[16,34].

## IV – Conclusion

We have used magnetic sedimentation to determine the size distribution of dispersed magnetic particles. The particles investigated here were of two kinds. We first studied different batches of single anionic and cationic nanoparticles of diameter D ~ 7 nm and polydispersity ranging from s = 0.2 to s = 0.4. Second, we dealt with polymer-nanoparticle hybrids obtained by electrostatic complexation. Different classes of polymers were utilized to this aim, such as polyelectrolytes and polyelectrolyte-neutral copolymers. The concentration profiles resulting from magnetic sedimentation were adjusted by the theoretical expression given in Eq. 9. In the model, we have made the assumption that the particles and hybrids are distributed in size according to a log-normal function (Eq. 7). However, other distributions can be envisaged. We have also noticed in the fitting procedure that the expression of the magnetic length scale λ (Eqs. 5 and 6) had to be slightly modified in order to reproduce the vibrating sample magnetometry data. The Langevin parameter L(ζ) where $\zeta(H) = \mu_0 \pi D^3 m_S H / 6 k_B T$ has been set to unity in Eq. 6, although its actual value was around 0.7. Using L(ζ) = 0.7 instead of 1 resulted in an increase of the median diameter $\tilde{D}_{Sed}$ by 10 % with respect to the data of Table I and no change in the polydispersity. This discrepancy could be due to systematic uncertainties in the determination of the parameters $\tilde{D}$, s and $m_S$, as obtained by one or the other technique. Measurements on very monodisperse nanoparticles should clarify this issue. When compared to dynamic light scattering, vibrating sample magnetometry and cryogenic transmission electron microscopy, magnetic sedimentation exhibits interesting properties : it allowed us to determine the state of aggregation of magnetic polymer-nanoparticle colloids. This technique could be applied to other composite systems loaded with magnetic particles, e.g. vesicles, minigels and endosomes[14,33].

**Acknowledgements** : We thank Jean-Claude Bacri, Andrej Cebers, Nicolas Schonbeck for fruitful discussions, Mikel Morvan from the Complex Fluids Laboratory in Bristol, Pennsylvania for the use of the block copolymers. Aude Michèle from the Laboratoire Liquides Ioniques et Interfaces Chargées at the Université Pierre et Marie Curie (Paris 6) is kindly acknowledged for the TEM and micro-diffraction experiments.

**Supporting Information Available:**
An electron beam microdiffraction experiment on S1 nanoparticles complexed with PTEA$_{5k}$-b-PAM$_{30k}$ is shown in this Section. Bragg reflections were observed and attributed to maghemite γ-Fe$_2$O$_3$ structure. This material is available free of charge via the Internet at http://pubs.acs.org